\documentclass[11pt]{article}
\def\half{\mbox{\small $\frac{1}{2}$}}

\def\vec#1{\ifmmode
\mathchoice{\mbox{\boldmath$\displaystyle\bf#1$}}
{\mbox{\boldmath$\textstyle\bf#1$}}
{\mbox{\boldmath$\scriptstyle\bf#1$}}
{\mbox{\boldmath$\scriptscriptstyle\bf#1$}}\else
{\mbox{\boldmath$\bf#1$}}\fi}

\usepackage{graphicx}% Include figure files
\usepackage{subfigure}
\usepackage{amsmath}
\usepackage{amssymb}
\usepackage{amsfonts}
\usepackage{booktabs,mathrsfs,verbatim}

\begin{document}

\begin{flushleft}
%\hspace{10 cm} Ofer Vitells \\
\hspace{10 cm} \today \\
\end{flushleft}
\vspace{0.1 cm}

\begin{center}
\Large{\bf A comment on estimating sensitivity to neutrino mass
hierarchy in neutrino experiments}
 \\
\normalsize
\vspace{0.5 cm}
\renewcommand{\thefootnote}{\fnsymbol{footnote}}
\begin{center}
Ofer Vitells$^{1,}$\footnote{ofer.vitells@weizmann.ac.il}, Alex
Read$^{2,}$\footnote{a.l.read@fys.uio.no}
\end{center}

\vspace{0.2 cm}

\noindent {\footnotesize $^1$ Department of Particle Physics,
Weizmann Institute of Science, Rehovot 76100, Israel\\$^2$
Department of Physics, University of Oslo, P.O.Box 1048 Blindern,
0316 Oslo, Norway\\ }

\renewcommand{\thefootnote}{\arabic{footnote}}
\setcounter{footnote}{0}

\vspace{0.2 cm}

\end{center}

\begin{abstract}
Recently it has been proposed, in the context of experiments
designed to resolve the neutrino mass hierarchy, to use the average
posterior probability of one of the hypotheses as a measure of
sensitivity of future experiments. This has led to sensitivity
estimates that are drastically lower than common conventions. We
point to the fact that such estimates can be severely misleading:
the probability that an experiment would actually produce a result
similar to the average value can be in fact negligibly small. We
emphasize again the simple relation between median significance and
the likelihood ratio evaluated with the ``Asimov'' data set, which
can be used to express experimental sensitivity in Bayesian terms as
well.
\end{abstract}

\section{Introduction}
One of the goals of future neutrino experiments is to resolve the
neutrino mass hierarchy, that is, determine the sign of $\Delta
m^2_{32}$. This is essentially a testing of two simple hypotheses,
since the absolute value $|\Delta m^2_{32}|$ is known to a high
level of accuracy~\cite{naka}. The hypotheses are denoted by NH
(normal hierarchy): $\Delta m^2_{32} > 0$ and IH (inverted
hierarchy):$\Delta m^2_{32} < 0$.

It was recently suggested to calculate the average value of the
posterior probability $P(\text{NH})$ under the hypothesis NH as a
measure of sensitivity of future experiments
\cite{XQian}\cite{XQian2}. This has led to sensitivity estimates
that are drastically lower than other common conventions. For
example, it was concluded that to reach a discovery sensitivity, an
experiment would need to have an average value of the test statistic
$\Delta\chi^2$ (which is equal to minus two times log of the
likelihood ratio) at an extraordinarily high level of $\sim 100$
\cite{ABBetal}. This conventionally corresponds to an
expected\footnote{Note that the term `expected' is often used
loosely in high energy physics in reference to either the mean or
the median. Here we will use the terms `average' or `expectation'
when referring to mathematical expectation.} significance of
$10\sigma$.

The general problem with calculating averages is that the result
strongly depends on the choice of the quantity which is being
averaged: the posterior probability calculated from the average
likelihood ratio, for example, is very different from the average
posterior probability, etc. This makes such quantities particularly
difficult to interpret. For probabilities in general, which are
confined to the range [0,1] and can have highly skewed
distributions, the average can be particularly misleading, since it
can represent highly unlikely outcomes. This is in fact already
quite clearly evident from Fig.~3 in Ref.~\cite{XQian}, which
compares the average posterior probability to its lower 90\%
quantile: for an experiment with an average $\Delta\chi^2$ of 40,
the 90\% quantile of $P(\text{IH})$ is  $10^{-5}$ while the average
is greater than $10^{-3}$. In other words, there is more than 90\%
probability that the experiment will produce a result much better
than its ``expectation''. It can also be seen that this discrepancy
is increasing with the average $\Delta\chi^2$, therefore the
probability of obtaining a result equal to or worse than the
expectation becomes exceedingly small.

In the following section we recall some of the asymptotic properties
of tests based on the likelihood ratio, in order to clarify the use
of the ``Asimov'' data set in determining conventional measures of
sensitivity, and the relation between them and the average posterior
probability. We use this to further illustrate the inappropriateness
of averaged probabilities as a measure of sensitivity.

%%----
%To see this more qualitatively, consider the case illustrated in
%Fig.1 of a test statistic which has a normal distribution and where
%the distance between the two hypotheses is $Z\sigma$. $Z$ is
%conventionally called the median or expected significance, or just
%the `sensitivity'. The \emph{median} of the posterior probability
%$P(H_0)=1-P(H_1)$ under $H_1$ is $(1+e^{Z^2/2})^{-1}$ , which for
%large $Z$ is comparable with the median $p$-value of $H_0$,
%$\Phi(-Z)$. The \emph{expected} posterior probability
%$\mathbb{E}[P(H_0)]$ however is larger than (see appendix A)
%$\half\Phi(-\frac{Z}{2})$, i.e. it is comparable to the median
%posterior probability of an experiment with only half of the
%expected significance. For $Z=10$, there is less than $3 \times
%10^{-4}$ probability of actually observing $P(H_0)$ larger than
%$\mathbb{E}[P(H_0)]$. Such experiment is thus almost guaranteed to
%achieve a much better result then the expectation.

\section{The conventional presentation of experimental sensitivity}

The median result that an experiment is expected to produce under a
given hypothesis is commonly used to present experimental
sensitivity, usually together with ``$\pm1\sigma$'' and
``$\pm2\sigma$'' bands, i.e.\ the corresponding quantiles. This has
been the main convention in high energy physics since at least the
days of LEP, see e.g. Refs.~\cite{delphi}\cite{cls}\cite{higgs}. For
the purpose of the following discussion we will assume that the
asymptotic distributions of the likelihood ratio test statistic
given in Ref.~\cite{asimov} are valid\footnote{This is essentially a
generalization of the approximation derived in Ref.~\cite{XQian} for
the distribution of $\Delta\chi^2$ under the assumption that the
data follow a gaussian distribution.}. Under those conditions, the
likelihood ratio test statistic

\begin{equation}
q = -2\log \frac{P(x|H_0)}{P(x|H_1)}
\end{equation}

\noindent is normally distributed. We denote the standard deviation
by $\sigma$ and the distance between the two hypotheses by
$Z\sigma$, as illustrated in Fig.~1.

\begin{figure}[ht!]
\begin{center}
\includegraphics*[width=8cm]{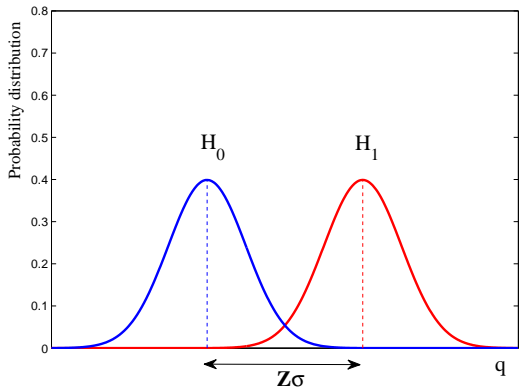}
\end{center}\caption{Distributions of likelihood ratio test statistic $q$ for the two hypotheses $H_0$ and $H_1$ and the definition of $Z$.}
\end{figure}

In frequentist terminology $Z$ is conventionally called the median
or expected significance. The median value of $q$ under hypothesis
$H_1$ is denoted by $q_A$. This corresponds to the value of the
likelihood ratio statistic obtained with the so called ``Asimov''
data set, which for poisson or gaussian data is just the expected
value of the data $x$ under hypothesis $H_1$. The \emph{observed
significance}, i.e.\ the distance between the observed $q$ and $H_0$
in units of standard deviation is given by

\begin{equation}
z^{obs} = \frac{q+q_A}{2\sqrt{q_A}}
\end{equation}

\noindent and the corresponding $p$-value is $1-\Phi(z^{obs})$,
where $\Phi$ is the standard normal cumulative distribution. The
\emph{median} significance which is obtained by $q=q_A$ is therefore

\begin{equation}
z^{med} = Z = \sqrt{q_A}.
\end{equation}

\noindent Note that while $q$ does \emph{not} have a $\chi^2$
distribution, the median significance \emph{is} related to the
median $q_A$ via the simple relation $Z = \sqrt{q_A}$.

%Furthermore although it is common to call $z^{obs}$ the
%`significance', the above results are by no means limited to
%frequentist interpretations.

The Bayesian posterior probability of $H_1$ (assuming equal prior
probabilities for both hypotheses) is given directly from the
likelihood ratio by

\begin{equation}
P(H_1) = (1 + e^{-q/2})^{-1}.
\end{equation}

\noindent Since $z^{obs}$ is a standard normal random variable, any
quantile of a monotonically related quantity such as $P(H_1)$ can be
immediately calculated by substituting the corresponding quantiles
of $z^{obs}$. For example, the central 68\% `sensitivity band' is
obtained by taking $z^{obs} = [Z+1,Z-1]$, that is

\begin{equation}
q_{\pm1} =q_A \pm 2\sqrt{q_A} = Z^2 \pm 2Z,
\end{equation}

\noindent and by substituting this into (4) one gets the
corresponding quantiles for $P(H_1)$.

The \emph{average} posterior probability is given by

\begin{equation}
E[P(H_1)] = \int_{-\infty}^{+\infty}dqf(q|H_1)(1 + e^{-q/2})^{-1}
\end{equation}

\noindent and by noting that $P(H_1)$ is smaller than $1/2$ when
$q<0$ the following simple bound can be derived:

\begin{eqnarray}
E[P(H_1)] & \leq & \half\int_{-\infty}^0dqf(q|H_1) +
\int_0^{+\infty}dqf(q|H_1) \\
& = & 1 - \half\Phi(-Z/2)\nonumber
\end{eqnarray}

\noindent which implies that the averaging has a similar effect,
roughly, to reducing $Z$ by half. In Fig.~2 we compare this bound on
the average value of $P(H_0)=1 - P(H_1)$ with several of its
quantiles from Eq.(4), which illustrates how the expectation is
pushed to the tail of the distribution as $q_A=Z^2$ increases. For
$q_A=100$, $E[P(H_0)]$ is above $10^{-7}$ (comparable to
``$5\sigma$''), which is well above the 1 per mil upper quantile.
The 1\% quantile for $q_A=100$ is already much lower at $10^{-12}$,
meaning that there is 99\% probability that the experiment will
produce a level of evidence at least as high. In other words, such
an experiment is almost guaranteed to produce an outcome with a
level of evidence that is vastly superior to its average value.

\begin{figure}[ht!]
\begin{center}
\includegraphics*[width=10cm]{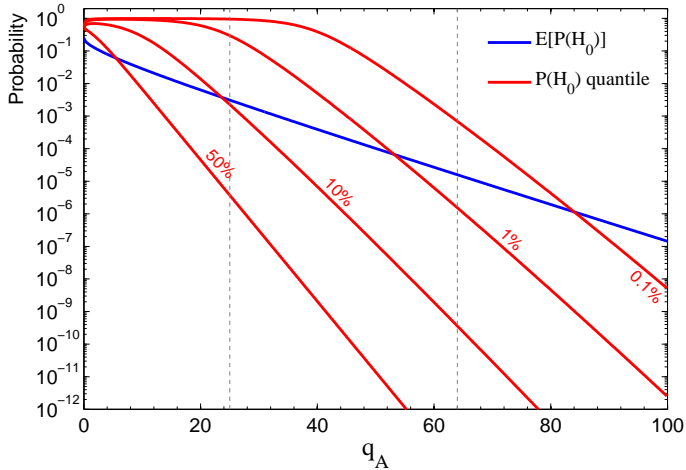}
\end{center}
\caption{Blue curve: lower bound on the average value of the
posterior probability $E[P(H_0)]$ under hypothesis $H_1$. The actual
average lies above the blue line. Red curve: several quantiles of
the distribution of $P(H_0)$ under hypothesis $H_1$. The two
vertical dotted lines correspond to Z=5 and Z=8.}
\end{figure}

We finally note that a similar problem will arise if one would
attempt to calculate the average $p$-value: this can be shown by
direct calculation to be equal to $1-\Phi(Z/\sqrt{2})$, which again
corresponds to a much lower significance level ($Z/\sqrt{2}$) than
the median.

\subsection{Another sensitivity measure}
A different sensitivity measure introduced in Ref.~\cite{XQian} and
applied e.g. in Ref.~\cite{Rubia}, was defined as ``the probability
of determining the correct hierarchy'', namely the probability that
the likelihood ratio will favor the true hypothesis, i.e.\ $P(q>0) =
1 - \Phi(Z/2)$. This obviously also leads to a sensitivity measure
that is equal to exactly half of the convention described above,
although for a very different reason. We therefore stress the very
different meanings of these two
definitions: \\
Determination of the mass hierarchy with 5$\sigma$ confidence level
(a ``5$\sigma$ discovery'') formally implies that the probability of
making an error, i.e.\ choosing the wrong hierarchy, is less than
$3\times10^{-7}$. Therefore:
\begin{itemize}
\item An experiment with $Z=5$ has a 50\% probability of making a
5$\sigma$ discovery, and a typical observed significance will be in
the range 4$\sigma$ -- 6$\sigma$ (with 68\% probability)

\item An experiment with $Z=10$ has a 100\% probability, i.e.\ is
\emph{guaranteed} to make a 5$\sigma$ discovery, and a typical
observed significance will be in the range  9$\sigma$ -- 11$\sigma$
(with 68\% probability). \end{itemize}

\noindent It should be clear that the definition of ``5$\sigma$
sensitivity'' adopted in \cite{Rubia} corresponds to the second case
above, i.e.\ to Z=10.

\section{Conclusions}
The average posterior probability can severely under-estimate the
actual sensitivity of an experiment, in terms of its probability to
achieve high levels of evidence. This can be seen by comparing the
average to the median and other quantiles that have a simple
relation to the median likelihood ratio test statistic evaluated
with the ``Asimov'' data set. Furthermore the sensitivity measure
that is defined by the ``probability of determining the correct
hierarchy'' leads to a similar effect. The meaning of such estimates
should be well understood and not confused with the common
convention.

\end{document}